\documentclass[preprint,12pt,authoryear]{elsart5p}

\usepackage{amsmath}
\usepackage{amssymb}
\usepackage{subfig}
\usepackage{graphics}
\usepackage{epsfig}

\journal{Physics Letters B}

\begin{document}
\renewcommand{\textfraction}{0.00000000001}
\renewcommand{\floatpagefraction}{1.0}
\topmargin -2.cm
\begin{frontmatter}
\title{First measurement of helicity-dependent cross sections in $\pi^0\eta$ 
photoproduction from quasi-free nucleons}


\author[basel]{A.~K{\"a}ser},
\author[basel]{M.~Dieterle},
\author[basel]{L.~Witthauer},
\author[basel]{S.~Abt},
\author[mainz]{P.~Achenbach},
\author[mainz]{P.~Adlarson},
\author[hiskp]{F.~Afzal},
\author[regina]{Z.~Ahmed},
\author[mainz]{J.~Ahrens},
\author[glasgow]{J.R.M.~Annand},
\author[mainz]{H.J.~Arends},
\author[edinburgh]{M.~Bashkanov},
\author[hiskp]{R.~Beck},
\author[mainz]{M.~Biroth},
\author[dubna]{N.S.~Borisov},
\author[pavia]{A.~Braghieri},
\author[gdwu]{W.J.~Briscoe},
\author[mainz]{F.~Cividini},
\author[halifax]{C.~Collicott},
\author[pavia]{S.~Costanza\thanksref{1}},
\author[mainz]{A.~Denig},
\author[mainz,gdwu]{E.J.~Downie},
\author[mainz]{P.~Drexler},
\author[tomsk]{A.~Fix},
\author[basel]{S.~Garni},
\author[glasgow,edinburgh]{D.I.~Glazier},
\author[dubna] {I.~Gorodnov},
\author[mainz]{W.~Gradl},
\author[basel]{M.~G\"unther},
\author[moscow]{G.M.~Gurevich},
\author[mainz]{L. Heijkenskj{\"o}ld},
\author[allison]{D.~Hornidge},
\author[regina]{G.M.~Huber},
\author[mainz,dubna]{V.L.~Kashevarov},
\author[edinburgh]{S.~Kay},
\author[basel]{I.~Keshelashvili\thanksref{2}},
\author[moscow]{R.~Kondratiev},
\author[zagreb]{M.~Korolija},
\author[basel]{B.~Krusche}\ead{bernd.krusche@unibas.ch},
\author[dubna]{A.B.~Lazarev},
\author[moscow]{V.~Lisin},
\author[glasgow]{K.~Livingston},
\author[basel]{S.~Lutterer},
\author[glasgow]{I.J.D.~MacGregor},
\author[kent]{D.M.~Manley},
\author[mainz,allison]{P.P.~Martel},
\author[giessen]{V.~Metag},
\author[bochum]{W.~Meyer},
\author[amherst]{R.~Miskimen},
\author[mainz]{E.~Mornacchi},
\author[moscow,amherst]{A.~Mushkarenkov},
\author[dubna]{A.B.~Neganov},
\author[mainz]{A.~Neiser},
\author[basel]{M.~Oberle},
\author[mainz]{M.~Ostrick},
\author[mainz]{P.B.~Otte},
\author[regina]{ D.~Paudyal},
\author[pavia]{P.~Pedroni},
\author[moscow]{A.~Polonski},
\author[mainz,ucla]{S.N.~Prakhov},
\author[bochum]{G.~Reicherz},
\author[jerusalem]{G.~Ron},
\author[basel]{T.~Rostomyan\thanksref{3}},
\author[halifax]{A.~Sarty},
\author[mainz]{C.~Sfienti},
\author[mainz,gdwu]{V.~Sokhoyan},
\author[hiskp]{K.~Spieker},
\author[mainz]{O.~Steffen},
\author[gdwu]{I.I.~Strakovsky},
\author[basel]{Th.~Strub},
\author[zagreb]{I.~Supek},
\author[hiskp]{A.~Thiel},
\author[mainz]{M.~Thiel},
\author[mainz]{A.~Thomas},
\author[mainz]{M.~Unverzagt},
\author[dubna]{Yu.A.~Usov},
\author[mainz]{S.~Wagner},
\author[basel]{N.K.~Walford},
\author[edinburgh]{D.P.~Watts},
\author[basel,glasgow]{D.~Werthm\"uller},
\author[mainz]{J.~Wettig},
\author[mainz]{M.~Wolfes},
\author[edinburgh]{L.~Zana}

\address[basel]{Department of Physics, University of Basel, Basel, Switzerland}
\address[mainz]{Institut f\"ur Kernphysik, University of Mainz, Mainz, Germany}
\address[hiskp]{Helmholtz-Institut f\"ur Strahlen- und Kernphysik, University of Bonn, Bonn, Germany}
\address[regina]{University of Regina, Regina, SK S4S 0A2 Canada}
\address[kent]{Kent State University, Kent, OH, USA}
\address[glasgow]{SUPA School of Physics and Astronomy, University of Glasgow, Glasgow, G12 8QQ, UK}
\address[edinburgh]{SUPA School of Physics, University of Edinburgh, Edinburgh EEH9 3JZ, UK}
\address[dubna]{Joint Institute for Nuclear Research,141980 Dubna, Russia}
\address[pavia]{INFN Sezione di Pavia, Pavia, Italy}
\address[gdwu]{Center for Nuclear Studies, The George Washington University, Washington, DC, USA}
\address[halifax]{Department of Astronomy and Physics, Saint Marys University, Halifax, Canada}
\address[tomsk]{Laboratory of Mathematical Physics, Tomsk Polytechnic University, Tomsk, Russia}
\address[moscow]{Institute for Nuclear Research, Moscow, Russia}
\address[allison]{Mount Allison University, Sackville, New Brunswick E4L 1E6, Canada}
\address[zagreb]{Rudjer Boskovic Institute, Zagreb, Croatia}
\address[giessen]{II. Physikalisches Institut, University of Giessen, Germany}
\address[bochum]{Institut f\"ur Experimentalphysik, Ruhr Universit\"at, 44780 Bochum, Germany}
\address[amherst]{University of Massachusetts Amherst, Amherst, Massachusetts 01003, USA}
\address[ucla]{University of California at Los Angeles, Los Angeles, CA, USA}
\address[jerusalem]{Racah Institute of Physics, Hebrew University of Jerusalem, Jerusalem 91904, Israel}

\thanks[1] {also at Dipartimento di Fisica, Universit\`a di Pavia, Pavia, Italy.}
\thanks[2] {Now at Institut f\"ur Kernphysik, FZ J\"ulich, 52425 J\"ulich, Germany}
\thanks[3] {Now at Department of Physics and Astronomy., Rutgers University, Piscataway, New Jersey, 08854-8019, USA}

\begin{abstract}
The helicity-dependent cross sections for the photoproduction of $\pi^0\eta$ pairs have been measured for the 
first time. The experiment was performed at the tagged photon facility of the Mainz MAMI accelerator 
with the combined Crystal Ball - TAPS calorimeter. The experiment used a polarized deuterated butanol 
target and a circularly polarized photon beam. This arrangement allowed the $\sigma_{1/2}$ (photon and target 
spin antiparallel) and $\sigma_{3/2}$ (parallel spins) components to be measured for quasi-free production 
of $\pi^0\eta$ pairs off protons and neutrons. The main finding is that the two helicity components contribute 
identically, within uncertainties, for both participant protons and neutrons. The absolute couplings for protons 
and neutrons are also identical. This means that nucleon resonances contributing to this reaction in the 
investigated energy range  have almost equal electromagnetic helicity couplings, $A_{1/2}^{n,p}$ and $A_{3/2}^{n,p}$. 
Identical couplings for protons and neutrons are typical for $\Delta$ resonances and identical 
$A_{1/2}$ and $A_{3/2}$ components are only possible for $J\geq 3/2$ states, which constrains possible contributions 
of nucleon resonances. 
\end{abstract}

\end{frontmatter}

\section{Introduction}

Excited states of the nucleon decay almost exclusively by the emission of mesons. Photoproduction
of mesons is one of the principal tools for studying the nucleon excitation spectrum, which 
is crucial for the understanding of the strong interaction in the non-perturbative regime. So far, the 
data base for such reactions is dominated by single-meson production reactions. Two-body final states, 
such as $\pi N$, $\eta N$,..., are still the backbone of most partial wave analyses, however, the progress
in accelerator and detector techniques over the last two decades now allows studies of multi-meson
production reactions with comparable statistical and systematic uncertainties. This advance has opened a 
new window on spectroscopy that provides access to new questions about the excitation 
spectrum of nucleons. The obvious nucleon resonances of interest are those states that, due to 
their internal structure, have only small branching ratios for direct decays to the nucleon ground 
state. Rather they decay predominantly in cascades involving at least one intermediate excited state. 
In the quark model, configurations with both oscillators excited are likely candidates for such patterns.
This is discussed in Ref. \cite{Sokhoyan_15} in context with the photoproduction of $\pi^0$ pairs
through excitations of high lying nucleon resonances. However, even for medium high excitations such cascade 
decays can be very interesting when they allow one to study the states in detail which dominate the 
respective decay chain. Particularly interesting are final states with neutral mesons for which non-resonant 
background contributions are small.

The investigation of final states with meson pairs is challenging. The formalism discussing all possible 
observables for the photoproduction of single pseudoscalar mesons was laid out by Barker, Donnachie, 
and Storrow \cite{Barker_75} and later Chiang and Tabakin \cite{Chiang_97} gave the final answer how 
many different observables have to be measured for a `complete' experiment. In this case, differential 
cross sections, the three single polarization observables corresponding to a linearly polarized photon 
beam ($\Sigma$), a transversely polarized target ($T$), the polarization of the recoil nucleon ($P$),
and several double polarization observables have to be measured as functions of two kinematic variables
(usually incident photon energy $E_{\gamma}$ or invariant mass $W$ and meson center-of-momentum (cm)
polar angle). In total eight observables, when combined in the right way, are sufficient. However, 
even this has not yet been achieved for the most prominent reaction channels like pion and $\eta$ 
photoproduction. 

The situation is more difficult for the production of meson pairs. This was discussed 
in detail by Roberts and Oed \cite{Roberts_05}. The measurement of eight observables as functions 
of five kinematic parameters fixes only the magnitude of the amplitudes and 15 observables would be 
necessary to extract the complex phases also. This is certainly not practical, in particular not for 
reactions with small cross sections as production of $\pi\eta$ pairs. However, for specific questions
already the measurement of one well chosen polarization observable can give useful additional 
information. For final states with meson pairs it is not always necessary to explore the full
three-body structure of the final state. Already the analysis as a quasi two-body final state,
for example $\gamma N\rightarrow N X$ ($X$ = $\pi\eta$) can give valuable insights. Different 
partitions of the final state are also possible. An example for such an analysis is given in 
\cite{Gutz_14} for the production of $\pi\eta$ pairs off the nucleon measured with a linearly
polarized photon beam. In this case, one can define polarization observables in analogy to 
\cite{Barker_75,Chiang_97} using the polar angle of the combined $X$ particle system. 

In spite of the complexity of photoproduction of meson pairs, double pion production has been 
intensively explored during the last decade. 
(see e.g. \cite{Sokhoyan_15,Thiel_15,Dieterle_15,Kashevarov_12,Zehr_12,Sarantsev_08,Thoma_08} 
and Ref. therein) but more recently the production of $\pi\eta$ pairs has also moved into the focus. 
This decay is more selective since the $\eta$ meson, due to its isoscalar nature, can only be
emitted in transitions between two isospin $I=1/2$ $N^{\star}$ resonances or between two $I=3/2$
$\Delta$ states. Cross-over decays between $N^{\star}$ and $\Delta$ states are not permitted. 
The data base for this reaction has grown rapidly during the last few years. Total cross sections, 
invariant-mass distributions, and some polarization observables, have been measured for the production 
of $\eta\pi^0$ pairs off protons at LNS in Sendai, Japan \cite{Nakabayashi_06}, GRAAL at ESRF in 
Grenoble, France \cite{Ajaka_08}, ELSA in Bonn, Germany \cite{Gutz_14,Horn_08a,Horn_08b,Gutz_08,Gutz_10}, 
and at MAMI in Mainz, Germany \cite{Kashevarov_09,Kashevarov_10,Annand_15,Sokhoyan_18} (see \cite{Krusche_15} 
for a summary). The isospin dependence of this reaction has been investigated at low incident 
photon energies ($E_{\gamma}<1.4$~GeV) with measurements of the $\gamma d\rightarrow np \pi^0\eta$,
$\gamma d\rightarrow nn \pi^+\eta$, $\gamma d\rightarrow pp \pi^-\eta$, and $\gamma d\rightarrow d\pi^0\eta$
reactions at MAMI \cite{Kaeser_15,Kaeser_16}. For the quasi-free reactions, recoil nucleons detected in
coincidence with the mesons were used to identify the final state.  

Prior analyses of data for $\gamma p\rightarrow\pi^0\eta p$ have suggested the dominance
of $\Delta$ excitations decaying via $\eta$ emission to the $\Delta(1232)$ state
\cite{Gutz_14,Horn_08b,Kashevarov_09,Kashevarov_10}. Some minor contributions were attributed to the
$R\rightarrow S_{11}(1535)\pi$ intermediate state ($R$ = any nucleon resonance), and at higher incident
photon energies, to the decay of the $a_0$ meson. The isospin dependence \cite{Kaeser_15,Kaeser_16} was in 
excellent agreement with the assumption of the reaction chain 
$\gamma N\rightarrow\Delta\eta\rightarrow N\pi\eta$. This means that the cross sections for the production
of the same charge type of pions (neutral or charged) for proton and neutron targets were identical within
uncertainties. The cross sections for the production of neutral pions were, for both types of incident 
nucleons, twice as large as for charged pions. The same relations also hold for the decay of a primarily
excited $\Delta$ resonance, via pion emission to an $N^{\star}$ state, with subsequent $\eta$ decay to the
nucleon ground state. However, invariant-mass distributions of the meson-nucleon pairs favor
the $\Delta(1232)\eta$ intermediate state \cite{Kaeser_16}. The isobar model analysis of Fix and coworkers
\cite{Fix_10} identified major contributions from the $D_{33}$ partial wave as the initial state 
($\Delta(1700)3/2^-$ and $\Delta(1940)3/2^-$ resonances). The comprehensive analysis of differential cross 
sections and polarization observables in \cite{Gutz_14} quotes not only branching ratios into $\Delta(1232)\eta$ 
for the $\Delta(1700)3/2^-$, $\Delta(1900)1/2^-$, $\Delta(1905)5/2^+$, $\Delta(1910)1/2^+$, 
$\Delta(1920)3/2^+$, and $\Delta(1940)3/2^-$ states, but also branching ratios into $N(1535)\pi$. 

The present paper reports the results for the first measurement of any double polarization observable
for this reaction. Measured was the observable $E$ (see definition below), and the decomposition of the 
cross section $\sigma$ into its helicity-1/2 and helicity-3/2 parts, $\sigma_{1/2}$ and $\sigma_{3/2}$. 
This is the first measurement of the helicity structure for photoproduction of $\pi^0\eta$ pairs 
from quasi-free protons and neutrons. Even for free protons this observable has not yet been studied.
It is measured with a longitudinally polarized target and a circularly polarized photon beam, 
where $\sigma_{3/2}$ corresponds to parallel target-nucleon and photon-beam spin orientation 
and $\sigma_{1/2}$ to the anti-parallel orientation. The electromagnetic excitation of nucleon resonances 
in the $S_{11}$, $S_{31}$ and $P_{11}$, $P_{31}$ partial waves can only contribute to the $\sigma_{1/2}$ part, 
while nucleon resonances with larger spins may contribute to both $\sigma_{1/2}$ and $\sigma_{3/2}$. The measurement 
of the helicity decomposition for the latter is sensitive to the relative contribution of the $A_{3/2}$ and 
$A_{1/2}$ electromagnetic amplitudes of resonance excitations. These are important properties of the 
structure of the excited nucleon states predicted e.g. by quark models. 

The two states, $\Delta(1700)3/2^-$ and $\Delta(1940)3/2^-$, suggested by several analyses of existing data 
as dominant in the reaction up to invariant masses of 1.9~GeV, are both listed in the {\it Review of Particle Physics}
(RPP) \cite{PDG_16} with similar $A_{3/2}$ and $A_{1/2}$ couplings. The RPP estimates for the Breit-Wigner photon-decay 
amplitudes for the $\Delta(1700)3/2^-$ are $A_{1/2}=A_{3/2}=140\pm 30$ (all values for photon couplings in units 
of $10^{-3}{\rm GeV}^{-1/2}$). The most recent results come from the Bonn-Gatchina (BnGa) coupled channel analysis 
for $\pi^0$ pairs \cite{Sokhoyan_15} and $\pi^0\eta$ pairs \cite{Gutz_14}. Both papers quote values of 
$A_{1/2}=165\pm 20$ and $A_{3/2}=170\pm 25$. Previous analyses listed in RPP \cite{PDG_16} differ significantly in 
absolute values (for example between 58 and 226 for $A_{1/2}$) and partly also in the $A_{3/2}/A_{1/2}$ ratio. 
The RPP lists for the $\Delta(1940)3/2^-$ state only results from the BnGa analysis \cite{Sokhoyan_15,Gutz_14}, 
which are $A_{1/2}=170^{+110}_{-80}$ and $A_{3/2}=150\pm 80$, so that in this case the uncertainty of the 
$A_{3/2}/A_{1/2}$ ratio is still large. All these results come mostly from coupled-channel analyses of data 
which are not directly sensitive to the $\sigma_{3/2}/\sigma_{1/2}$ ratio (but only rather indirectly sensitive
via angular distributions etc.). The present experiment provides the first direct measurement of this ratio.   

\section{Polarization observable $E$ and helicity dependent cross sections $\sigma_{1/2}$ and $\sigma_{3/2}$}

The polarization observable $E$ and the helicity-dependent cross sections $\sigma_{1/2}$ and $\sigma_{3/2}$
can be measured with a circularly polarized photon beam of polarization $P_{\odot}$ and a longitudinally polarized 
target of polarization $P_{T}$. The cross sections $\sigma_{1/2}$ and $\sigma_{3/2}$ correspond to the antiparallel
$(\uparrow\downarrow)$ or parallel $(\uparrow\uparrow)$ configurations  of incident nucleon and photon spin
(details of the spin-helicity configurations are for example given in \cite{Drechsel_04}). In the full three-body 
formalism of \cite{Roberts_05} this would be the observable $P_z^{\odot}$. However, since we analyze only the 
fully integrated asymmetry, for which the definition is identical to the analysis of a two-body final state, 
as a short-hand notation we use `$E$' as in \cite{Barker_75}. The asymmetry and the two partial cross sections 
are then defined by 
\begin{equation}
E=\frac{\sigma_{1/2}-\sigma_{3/2}}{\sigma_{1/2}+\sigma_{3/2}}=
\frac{1}{P_{\odot}P_{T}}\cdot\frac{N_{1/2}-N_{3/2}}{(N_{1/2}-N_B)+(N_{3/2}-N_B)}~.
\label{eq:e}
\end{equation}
The right-hand side of the equation with the count rates $N_{1/2}$ and $N_{3/2}$, measured for the two spin 
configurations, ensures that all absolute normalizations (target density, beam flux, detection efficiencies,...) 
cancel in the count-rate ratio. Since molecular hydrogen cannot be polarized, solid deuterated butanol 
(${\rm C}_{4}{\rm D}_{10}{\rm O}$) was used as the target material. Therefore, a background count rate $N_B$ from 
reactions with nucleons bound in the unpolarized $J=0$ carbon and oxygen nuclei must be subtracted in the 
denominator. This background cancels in the cross section difference in the numerator.

There are different strategies for extracting the asymmetry $E$ and $\sigma_{1/2}$ and $\sigma_{3/2}$ from a measurement
with a butanol target. For the asymmetry $E$, in one approach, the denominator ($\sigma_{1/2}+\sigma_{3/2})$ was 
replaced by the results from a measurement of the unpolarized cross section $\sigma_0$ with a liquid deuterium target 
using $2\sigma_{0} = \sigma_{1/2}+\sigma_{3/2}$. This method needs absolutely normalized cross-section data for numerator 
and denominator. These results are labeled (A). For the second method, a measurement was performed to determine the background 
rate $N_B$ with a carbon foam target that had the same mass, volume, and density as the non-deuterium components 
of the butanol target. A small correction had to be applied for nucleons bound in oxygen nuclei because nuclear cross 
sections scale as $A^{2/3}$ rather than $A$. This method required only that the count rates measured 
with the butanol and the carbon foam were normalized to the beam flux. The results from this analysis are labeled (B). 
The systematic uncertainties of the two methods have different sources. However, the statistical uncertainties are highly 
correlated because they are dominated by the fluctuations of the small numerator, which is identical in both analyses.

For the cross sections $\sigma_{1/2}$ and $\sigma_{3/2}$, three different ways of extraction were explored. They are all
based on the relations
\begin{equation}
\begin{aligned}
\sigma_{1/2} &= \sigma_0\cdot(1+E)\\
\sigma_{3/2} &= \sigma_0\cdot(1-E)~,
\end{aligned}
\label{eq:s1}
\end{equation}
but different results were used for $E$ and $\sigma_0$.
\begin{itemize}
\item[$\bullet$]{Version (1):} $E$ was taken from analysis (A) (denominator from measurement with liquid 
deuterium target) and $\sigma_{0}$ also from the measurement with the unpolarized target.
\item[$\bullet$]{Version (2):} $E$ was taken from analysis (B) (carbon subtraction), but $\sigma_0$ again from the
liquid deuterium target. 
\item[$\bullet$]{Version (3):} $E$ and $\sigma_{0}$ were taken from carbon subtracted butanol data. 
\end{itemize}

\section{Complications for quasi-free production off bound nucleons}

Measurements for nucleons bound in light nuclei, which are necessary for neutrons, introduce some complications.
A trivial experimental one is the requirement to detect the recoil nucleons, which for neutrons in particular,
reduces the detection efficiency.

Nuclear Fermi motion smears structures in excitation functions and angular distributions. However, this
problem can be partly avoided by a complete reconstruction of the kinematics of the final state. 
For photoproduction off the deuteron, the final state is completely determined kinematically, within experimental 
resolution, when the four momenta of all produced mesons and the three-momentum direction of the recoiling
nucleon are measured \cite{Krusche_11}. In this case, only the kinetic energy of the recoil nucleon and the three 
momentum of the spectator nucleon (four parameters) are missing. These parameters can be reconstructed from 
energy and momentum conservation (four equations). This analysis determines the `true' center-of-momentum (cm) 
energy $W=\sqrt{s}$ in the incident-photon - participant-nucleon system. The obtained resolution for $W$ is poorer 
than that obtained with measurements with free proton targets for which $W$ can be directly reconstructed from the 
incident photon energy, but this is not a problem for smoothly varying cross sections. All values for $W$ used 
in this analysis have been reconstructed using this method.

More problematic are effects of final-state interactions (FSI) between the nucleons, between mesons and nucleons, 
and, for multi-meson production reactions, also between mesons. The comparison of cross sections measured for the 
photoproduction of $\pi^0\eta$ pairs off free protons and off bound protons \cite{Kaeser_15,Kaeser_16} shows that
such effects are relevant for this final state. On an absolute scale, free and quasi-free cross sections deviate 
on average by $\approx$30\%. Model calculations of such effects are difficult and not far advanced. Recently, some
results were published for estimates of FSI effects for differential cross sections in single pion and $\eta$ 
production \cite{Tarasov_16,Nakamura_17}. They were, however, not yet precise enough for numerical corrections
of measured data, clearly more efforts in theory are needed. Similar effects in the production of meson pairs 
and for polarization observables are almost unexplored in model calculations. However, there are some interesting 
experimental results from the comparison of photoproduction reactions off free protons and quasi-free protons bound 
in the deuteron. As mentioned above, differential cross sections for $\eta\pi$ production are significantly
different for free and quasi-free protons, cross sections for single $\pi^0$ production are even more different
in some energy regions \cite{Dieterle_14,Dieterle_18}, also the results for pion pairs differ up to 20\%, 
while effects in $\eta$ production are insignificant. This means that FSI effects on absolute cross sections 
are strongly reaction dependent. However, polarization observables seem to be effected in a completely different 
way. We have previously tested this for the polarization observable $I^{\odot}$ measured with a circularly polarized 
photon beam on unpolarized target for the production of meson pairs \cite{Kaeser_16,Oberle_13,Oberle_14}. 
No significant effects were found for $\pi\pi$ and $\pi\eta$ pairs. Also, for the helicity asymmetry $E$, 
as defined in this paper, no effects were found for single $\pi^0$ \cite{Dieterle_17} and $\eta$ production 
\cite{Witthauer_16,Witthauer_17,Witthauer_17a}, although the effects on absolute cross sections for
$\pi^0$ production are substantial. For $\pi\eta$ production also shapes of invariant mass distributions
of meson-meson and meson-nucleon pairs are basically unaffected \cite{Kaeser_16}.  
      
Therefore, one expects significant FSI effects for the absolute scale of the $\sigma_{1/2}$ 
and $\sigma_{3/2}$ cross sections, but only minor effects for $E$ and the $\sigma_{3/2}/\sigma_{1/2}$ cross-section 
ratio. Unfortunately, since double polarization data of this type are not yet available for free protons, 
this cannot be tested directly. 

\section{Experimental Setup}
The present results are based on the same data set that was previously used to extract the polarization 
observable $E$ for the production of $\eta$ mesons \cite{Witthauer_16,Witthauer_17} and for $\pi^0$ 
mesons \cite{Dieterle_17} off quasi-free nucleons (most details are given in \cite{Witthauer_17}). 
Therefore, we give only a short summary of the experimental details.

The measurements were performed at the electron accelerator MAMI in Mainz, Germany \cite{Kaiser_08}. The electron 
source was an optically pumped gallium-arsenide-phosphor (GaAsP) photocathode \cite{Aulenbacher_97} delivering 
polarized electrons.
During four beam times, which were analyzed for the present results, the electrons were accelerated to 
energies close to 1.6~GeV in the accelerator stages of MAMI. The high energy electrons produced 
bremsstrahlung in a Co-Fe alloy (Vacoflux50, 10 $\mu$m thickness) and in this process the longitudinally 
polarization of the electrons was transferred to circular polarization of the photons according to \cite{Olsen_59}:
\begin{equation}
P_{\gamma}=P_{e^{-}}\cdot\frac{4x-x^{2}}{4-4x+3x^{2}}~,\label{eq:olsen}
\end{equation}
where $P_{e^{-}}$ and $P_{\gamma}$ are the degrees of polarization of the electrons and the photons, respectively, and
$x=E_{\gamma}/E_{e^-}$. The electron polarization was measured periodically with a Mott polarimeter close to the 
electron source and monitored  with a M$\o$ller polarimeter viewing the ferromagnetic bremsstrahlung foil. 
Both results were in good agreement and the average electron polarization was $P_{e^{-}}\approx 83\%$. 

The photon beam was energy tagged with the Glasgow spectrometer \cite{McGeorge_08} with a typical resolution of 4~MeV,
which results from the widths of the 353 plastic scintillators used in the focal plane detector for detection of the
post-bremsstrahlung electrons. This detector covers 5 - 93\% of the incident electron energies but the part 
corresponding to high electron energies (low photon energies) was deactivated to increase counting statistics 
for high-energy photons. The active photon-energy range spanned from $E_{\gamma}\approx400$~MeV to 1450~MeV. 

The photon beam size was defined by a collimator with 2~mm diameter, producing a beam-spot diameter of 9~mm 
on the production target. The longitudinally polarized target had a diameter of 19.8~mm and a length of 20~mm.
It consisted of butanol beads of average diameter $\approx$1.9~mm \cite{Rohlof_04}. The deuterons in the butanol 
molecules were polarized by Dynamic Nuclear Polarization (DNP) \cite{Bradtke_99} in a strong magnetic field (1.5~T) 
at a temperature of 25 mK. After the target had been polarized, the polarizing magnet was replaced by a small 
solenoidal holding coil with a magnetic field of 0.6 T. Relaxation times of more than 2000 h and polarization 
degrees around 60\% were achieved. 

However, as discussed in detail in \cite{Witthauer_16,Witthauer_17} the target polarization was not homogeneous 
across the target diameter for the first three beam times. Thus, the standard NMR measurements of the 
polarization did not correspond to the effective polarization in the target center hit by the photon beam.
The problem was due to small inhomogeneities of the field of the 1.5~T magnet, used for the DNP process, 
combined with the very narrow NMR resonance of deuterated butanol doped with trityl-radicals. The fourth 
beam time used a less sensitive radical (Tempo), which resulted in lower polarization (55\%) that, however, 
could be determined much more accurately. The polarization of all beam times were then recalibrated to the 
last beam time using the measured asymmetries for the photoproduction of $\eta$-mesons.   

Furthermore, the NMR measurements determine only the polarization of the deuterium nuclei. For the effective 
polarization of the nucleons bound in the deuteron one must take into account the $d$-wave component in the 
deuteron wave function. This results in a downward correction of the measured polarization degrees on the 
order of 8\% \cite{Rondon_99}.

The experimental setup, combining the Crystal Ball and TAPS detectors with additional devices for charged 
particle identification, was identical to the one used for the results reported in 
\cite{Witthauer_16,Witthauer_17,Dieterle_17}, which used the same data set. The electromagnetic calorimeter
combined the Crystal Ball (CB) \cite{Starostin_01} and TAPS \cite{Gabler_94} detectors. The first 
(672 NaI(Tl) crystals) covered the solid angle for polar angles between $20^{\circ}$ and $160^{\circ}$
and the second (384 BaF$_{2}$ crystals) covered as a forward wall polar angles between $5^{\circ}$ and 
$21^{\circ}$. A cylindrically shaped charged-particle identification detector (PID) \cite{Watts_05}, 
consisting of 24 plastic scintillators, was mounted inside the CB around the target and a 5~mm thick 
plastic scintillator was mounted in front of each BaF$_{2}$ crystal for charged particle identification 
(CPV detector).

\begin{figure*}[!thb]
\centerline{
\resizebox{0.5\textwidth}{!}{\includegraphics{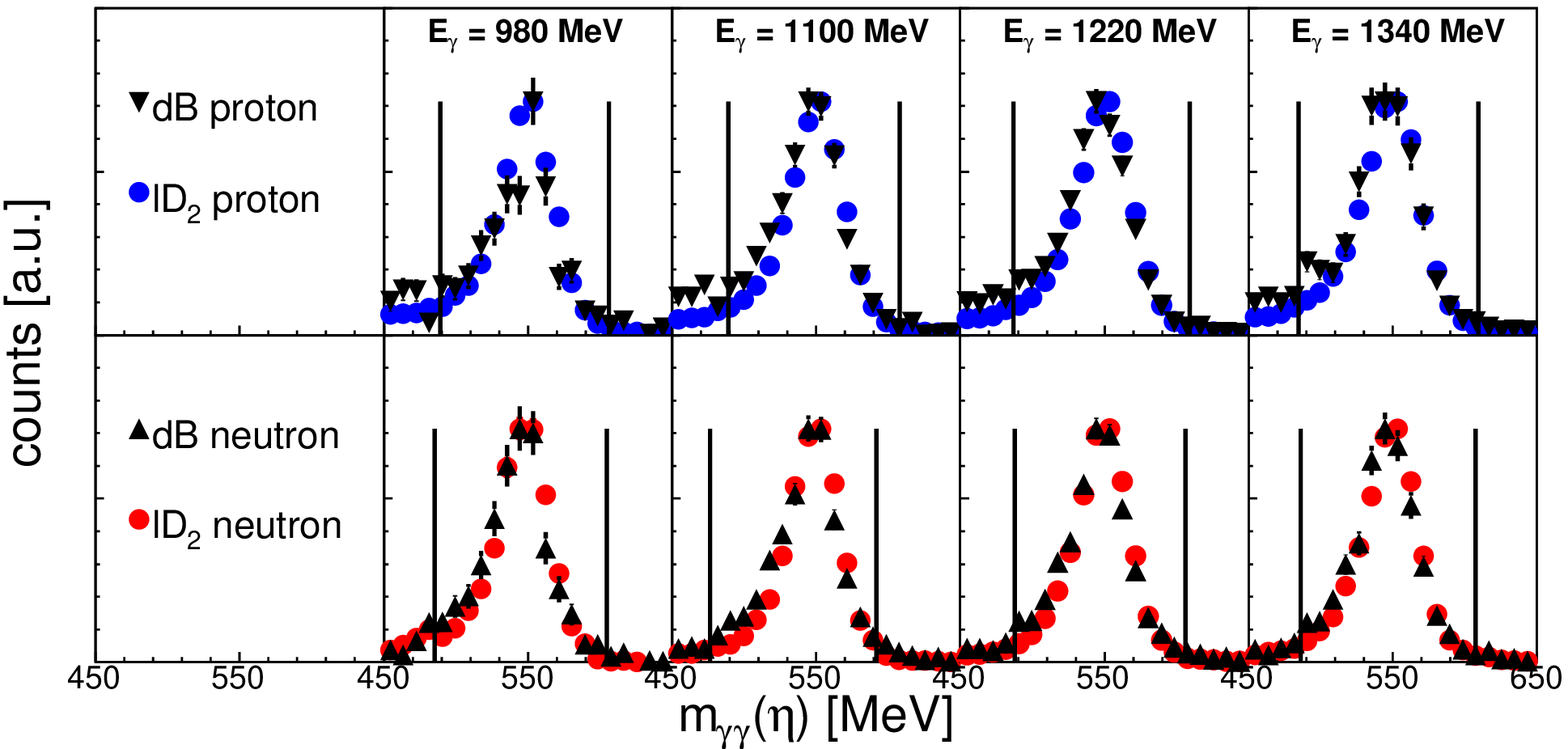}}
\resizebox{0.5\textwidth}{!}{\includegraphics{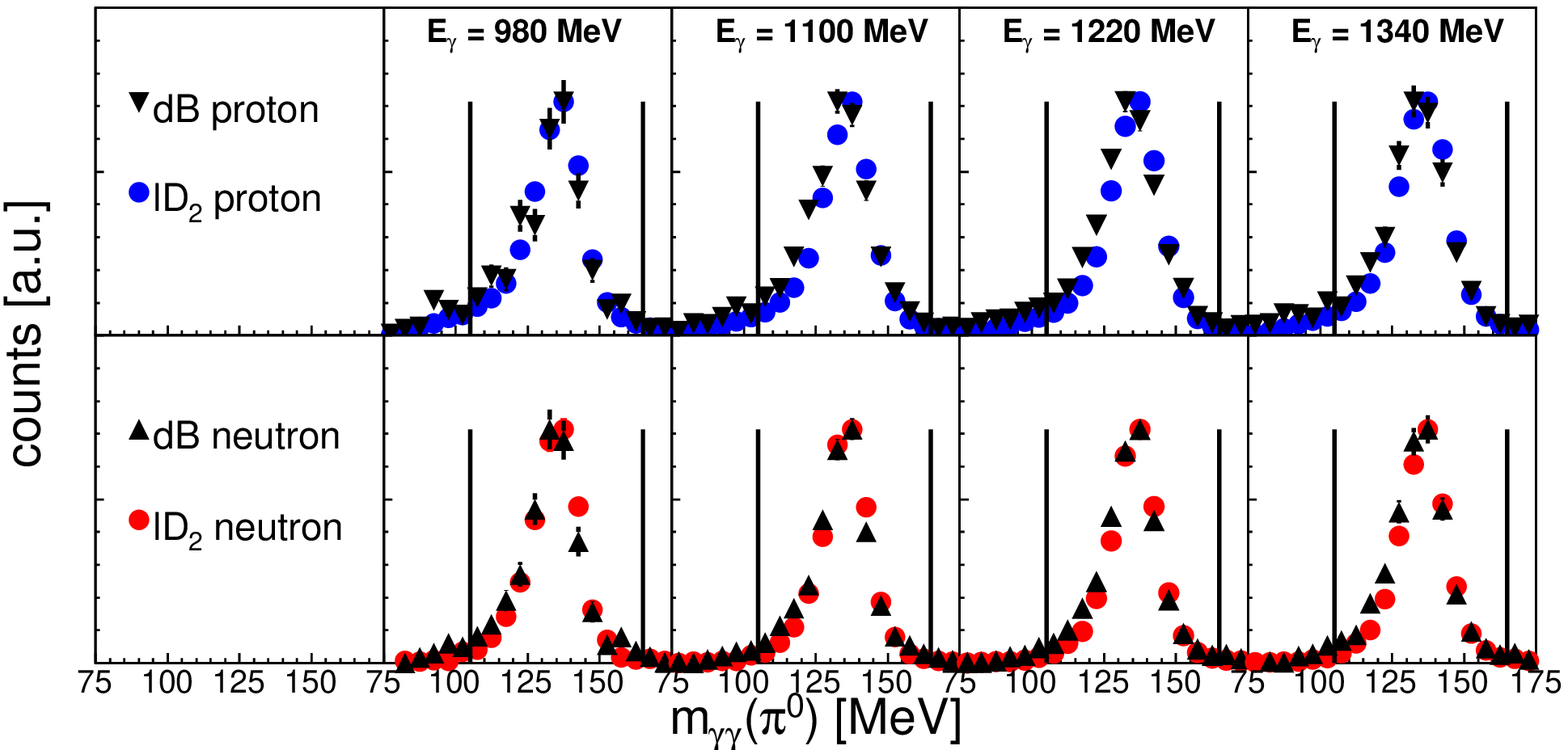}}}
\caption{Left-hand side: Invariant-mass distributions of the photon pairs assigned to decays of the 
$\eta$ mesons. Upper row: coincidence with recoil protons, lower row: coincidence with recoil 
neutrons. Centers of energy bins at 980~MeV, 1100~MeV, 1220~MeV, and 1340 MeV. 
Vertical lines indicate the experimental cuts. Filled spheres: results for liquid deuterium target,
black triangles: butanol target.
Right-hand side: Invariant-mass distributions of the photon pairs assigned to $\pi^0$ decays.  
Notation as for left-hand side.}
\label{fig:invmass}
\end{figure*}

The experimental trigger was based on a hit-cluster multiplicity condition for hits in the combined calorimeter 
and a sum threshold for the total energy deposition in the CB. Only events with at least two cluster hits in
the calorimeter were selected. Since each hit activates an a priori unknown number of detector modules, this 
condition was approximately imposed by dividing TAPS azimuthal coverage into six equal triangular sectors 
and the CB into sectors of 16 adjacent modules. Only events that activated at least two sectors 
were accepted. Furthermore, it was required that the analog sum of the energy signals from the CB 
exceeded 250~MeV. This condition removed a large fraction of electromagnetic background in the calorimeter. 
Events from single $\pi^0$ decays, with both photons in TAPS, were thus not included in the trigger, but 
this is irrelevant for the $\pi^0\eta$ final state. In the offline analysis, event-selection conditions
considered only meson-decay photons. This avoided systematic uncertainties from the unpredictable energy 
deposition of recoil neutrons in the calorimeter.

\section{Data Analysis}

The data analysis was based on the methods developed for the reaction identification of meson pairs
($\pi\pi$ and $\pi\eta$) described in detail in \cite{Dieterle_15,Kaeser_15,Kaeser_16,Oberle_13,Oberle_14}
for measurements with unpolarized liquid deuterium targets. The treatment of the unpolarized background
from `heavy' nuclei (carbon, oxygen), present in the butanol target, is described in 
\cite{Witthauer_16,Witthauer_17,Dieterle_17} for photoproduction of $\eta$ and $\pi^0$ mesons. For the latter, 
in addition to the data from the polarized butanol target, measurements with a carbon foam target and a 
liquid deuterium target were analyzed.

All detector modules were calibrated for their energy and timing response as discussed in detail in 
\cite{Dieterle_18}. Background from random coincidences with the tagger was subtracted as discussed 
in \cite{Werthmueller_14}. In the first step of particle identification, hits in the calorimeter were 
classified as `charged' or `neutral' depending on the response of the PID and the CPV. Subsequently, 
for hits in TAPS, pulse-shape analysis (PSA) and time-of-flight versus energy analysis were used for
the separation of photons from protons and neutrons as in \cite{Kaeser_16,Dieterle_18}. The only 
remaining ambiguity was that photons and neutrons in the CB cannot be distinguished event-by-event 
(see e.g. \cite{Werthmueller_14,Witthauer_13}) by the detector response. The timing resolution is only
modest due to the short time-of-flight distance, PSA methods cannot be applied, and cluster-size 
distributions discriminate not on an event-to-event basis. 

Events accepted for further analysis were those with four neutral and one charged hit for the 
$\gamma d\rightarrow \pi^0\eta p(n)$ reaction and five neutral hits for the 
$\gamma d\rightarrow \pi^0\eta n(p)$ reaction (nucleons in parentheses are undetected spectators). 

As discussed in detail in \cite{Kaeser_16}, neutral hits in the CB were assigned to photons or neutrons
using a $\chi^2$ analysis. For events with four or five neutral hits, the invariant masses of 
all possible pair combinations were compared to the $\pi^0$ and $\eta$ masses. 
The $\chi^2$ was defined by
\begin{equation}
\label{eq:chi2}
\chi^{2}(k) = \sum_{i=1}^{2}\left 
(\frac{m_{\pi^0,\eta}-m_{i,k}}{\Delta m_{i,k}}\right)^{2} ~~{\rm with}~~ k=1,..,n_p,
\end{equation} 
where the $m_{i,k}$ are the invariant masses of the $i$-th pair in the $k$-th permutation of the hits 
and $\Delta m_{i,k}$ is the corresponding uncertainty from the experimental energy and angular resolution. 
Both were computed event-by-event. For events with exactly four neutral hits, this analysis was used only 
to find the most probable combination of the four decay photons relating to a parent $\pi^0$ and $\eta$. 
For events with five neutral hits, the remaining hit was assigned to the neutron. In order to suppress 
combinatorial background, the hypothesis of $\pi^0\pi^0$ pairs was also tested and the event was discarded 
when such a combination resulted in a smaller $\chi^2$ than any of the $\pi^0\eta$ hypotheses. 
One-dimensional projections of the two-dimensional $\eta - \pi$ invariant mass spectra are shown in 
Fig.~\ref{fig:invmass}. As expected, the results for the liquid deuterium and the solid butanol targets 
are practically identical, in spite of the heavy-nuclei background in the butanol spectra, because Fermi 
motion does not influence invariant masses.

\begin{figure}[!htb]
\centerline{
\resizebox{0.5\textwidth}{!}{\includegraphics{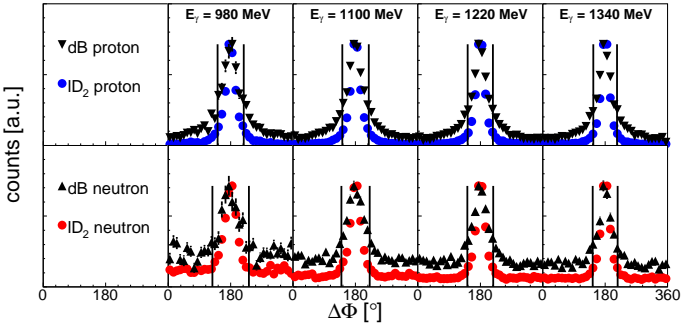}}}
\caption{Coplanarity spectra. Upper row: coincidence with recoil protons, lower row: coincidence 
with recoil neutrons for the same energy ranges as Fig.~\ref{fig:invmass} and with same notation.}
\label{fig:cop}
\end{figure}

\begin{figure*}[!thb]
\centerline{
\resizebox{0.8\textwidth}{!}{\includegraphics{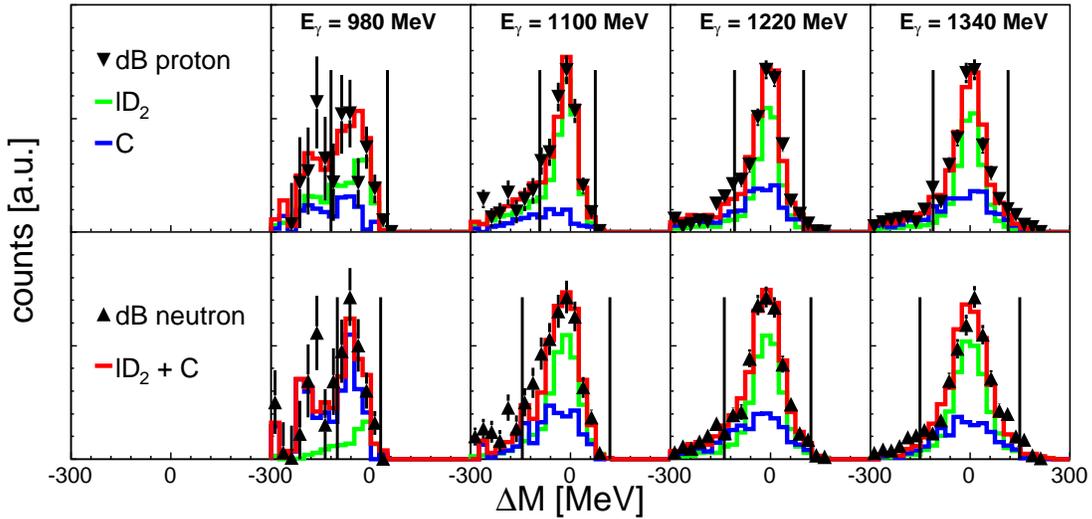}}}
\caption{Missing-mass spectra extracted from Eq.~\ref{eq:mm}. Black triangles: butanol target, 
green histograms: liquid deuterium target, blue histograms: carbon foam target, red histograms:
sum of liquid deuterium and carbon foam. Upper row: coincidence with recoil protons, 
lower row: coincidence with recoil neutrons. Vertical lines: analysis cuts.}
\label{fig:mismas}
\end{figure*}

Background not eliminated by the invariant mass analysis can arise for the liquid deuterium and butanol 
targets from photoproduction of $\pi^0$ pairs or other reactions with multi-photon final states. 
For the butanol target, background can also arise from reactions with the nucleons bound in the unpolarized 
nuclei. Due to the larger Fermi momenta in heavier nuclei, this leads to a larger width of the signal. 
Such backgrounds can be removed by analyses of the reaction kinematics. The most basic condition is the 
coplanarity of the two mesons and the recoil nucleon. Due to momentum conservation, the difference in the 
azimuthal angle $\Delta\Phi$ between the $\eta\pi^0$ pair and the recoil nucleon must be 180$^{\circ}$. 
This is normally not the case when additional particles have escaped detection or four photons have 
been wrongly assigned to the decay of a $\pi^0$ and an $\eta$ meson. The coplanarity spectra shown 
in Fig.~\ref{fig:cop} show a clear peak at $180^{\circ}$ and only events between $\pm 36^{\circ}$ around 
the peak were accepted. 

Even more powerful is the analysis of the missing mass. For this analysis, the recoil nucleon was treated 
as a missing particle (although it was detected) and its mass was calculated from the four momenta of the 
incident photon $P_{\gamma}$, the initial-state nucleon $P_{N}$, the final-state pion $P_{\pi^0}$, and the 
$\eta$ meson $P_{\eta}$:  
\begin{equation}
\Delta M = \left|P_{\gamma}+P_{N}-P_{\pi}-P_{\eta}\right| -m_N\ ,
\label{eq:mm}
\end{equation} 
where the nucleon mass $m_{N}$ was subtracted so that true $\gamma N\rightarrow N\pi^0\eta$ events were 
expected at $\Delta M = 0$. In Eq.~\ref{eq:mm}, $P_{N}$ is unknown due to the contribution of the Fermi
momentum to the four momentum. The Fermi momentum was set to zero. This results in a broadening of the
$\Delta M$ distribution which is more important for reactions with nucleons bound in the heavier nuclei 
than for the nucleons from the deuteron. This analysis was done for the butanol, the liquid deuterium, 
and the carbon foam target. As discussed in \cite{Witthauer_16,Witthauer_17,Dieterle_17} for other 
final states, the spectra from all three targets were normalized absolutely on the basis of photon flux, 
target density, etc. and are compared in Fig.~\ref{fig:mismas}. The sum of the liquid deuterium and 
carbon data agree well with the butanol data so that the contribution of reactions on quasi-free nucleons 
in deuterium can be precisely determined for the measurement with the butanol target. This was only 
important for analysis (B) of the asymmetry $E$, for which the denominator was taken from the measurement 
with the butanol target. For analysis (A), only the difference between the two helicity states in the 
numerator was used, for which the unpolarized carbon background cancels and the denominator was taken 
directly from the measurement with the liquid deuterium target. 

The measured yields were normalized absolutely with respect to the incident photon flux, the target 
density, the $\pi^0$ and $\eta$ decay branching ratios into two photons \cite{PDG_16}, and the 
detection efficiency. The detection efficiency was determined with Monte Carlo (MC) simulations which 
employ the Geant4 \cite{Geant4} tool kit. As discussed in \cite{Kaeser_16}, the generator of 
reaction-kinematics input into the simulation was based on the dominant 
$\Delta^{\star}\rightarrow \Delta(1232)\eta\rightarrow N\eta\pi^0$ decay chain. It included the effects 
from the Fermi motion of the bound nucleons using the Paris-potential parameterization for the deuteron 
wave function in momentum space \cite{Lacombe_81}. However, the effects of Fermi motion were mostly 
eliminated by the kinematic reconstruction of the final state. The MC simulations are precise and 
reliable for the detection of photons. There were, however, imperfections in the MC for 
the detection of recoil nucleons where these particles were emitted into the transition 
area between CB and TAPS. There inert materials from support structures, cables, etc. were not 
implemented in the MC with sufficient accuracy. Therefore, as in \cite{Dieterle_17,Werthmueller_14}, 
corrections based on the analysis of reactions such as 
$\gamma p\rightarrow p\eta$ and $\gamma p\rightarrow n\pi^0\pi^+$ measured with a liquid hydrogen target 
were applied. However, they mostly cancel in the asymmetries.    

All results are given as a function of the reconstructed invariant mass $W$ defined as:
\begin{equation}
W=\sqrt{s} = \left|P_{\pi}+P_{\eta}+P_{N}\right|,
\label{eq:W}
\end{equation}
where $P_{\pi}$, $P_{\eta}$, and $P_N$, are the four momenta of the $\pi^0$, the $\eta$, and the recoil
nucleon, respectively. The four momenta of the pion and the eta were obtained from the decay photons 
measured in the calorimeter, while the four momentum of the recoil nucleon was defined by its measured 
azimuthal and polar angles as well as overall momentum and energy conservation 
(see e.g. \cite{Kaeser_16,Krusche_11,Witthauer_16,Werthmueller_14}). 

The systematic uncertainty of the asymmetry is dominated by the uncertainty of the polarization  
of the photon beam (2.7\%) and the target (10\%) \cite{Witthauer_17}. The latter was very conservatively 
estimated due to the necessary recalibration of the first three beam times. Most other uncertainties 
cancel in the ratio of Eq.~\ref{eq:e}. Only higher-order effects from either the normalization to data 
from the measurement with a liquid deuterium target or the subtraction of the carbon background in the 
denominator can contribute. This means that only the difference between systematic effects for the three 
target types matters. Consequently, photon and recoil nucleon detection efficiencies and kinematic cuts 
are much less important than for absolute cross-section measurements. Systematic effects are further 
reduced in the comparison of the asymmetry for recoil protons and neutrons.

The $\sigma_{1/2}$ and $\sigma_{3/2}$ cross sections also carry the uncertainty from the absolute 
normalization (photon flux, target density), estimated to be between 5\% - 7\% 
\cite{Kaeser_15,Kaeser_16,Dieterle_17}, and uncertainties from the MC simulations of detector acceptance 
estimated in the range 5\% - 10\%. However, these uncertainties largely cancel in the comparison of the 
two helicity cross sections. 
 
\section{Results}

\begin{figure*}[!thb]
\centerline{
\resizebox{0.5\textwidth}{!}{\includegraphics{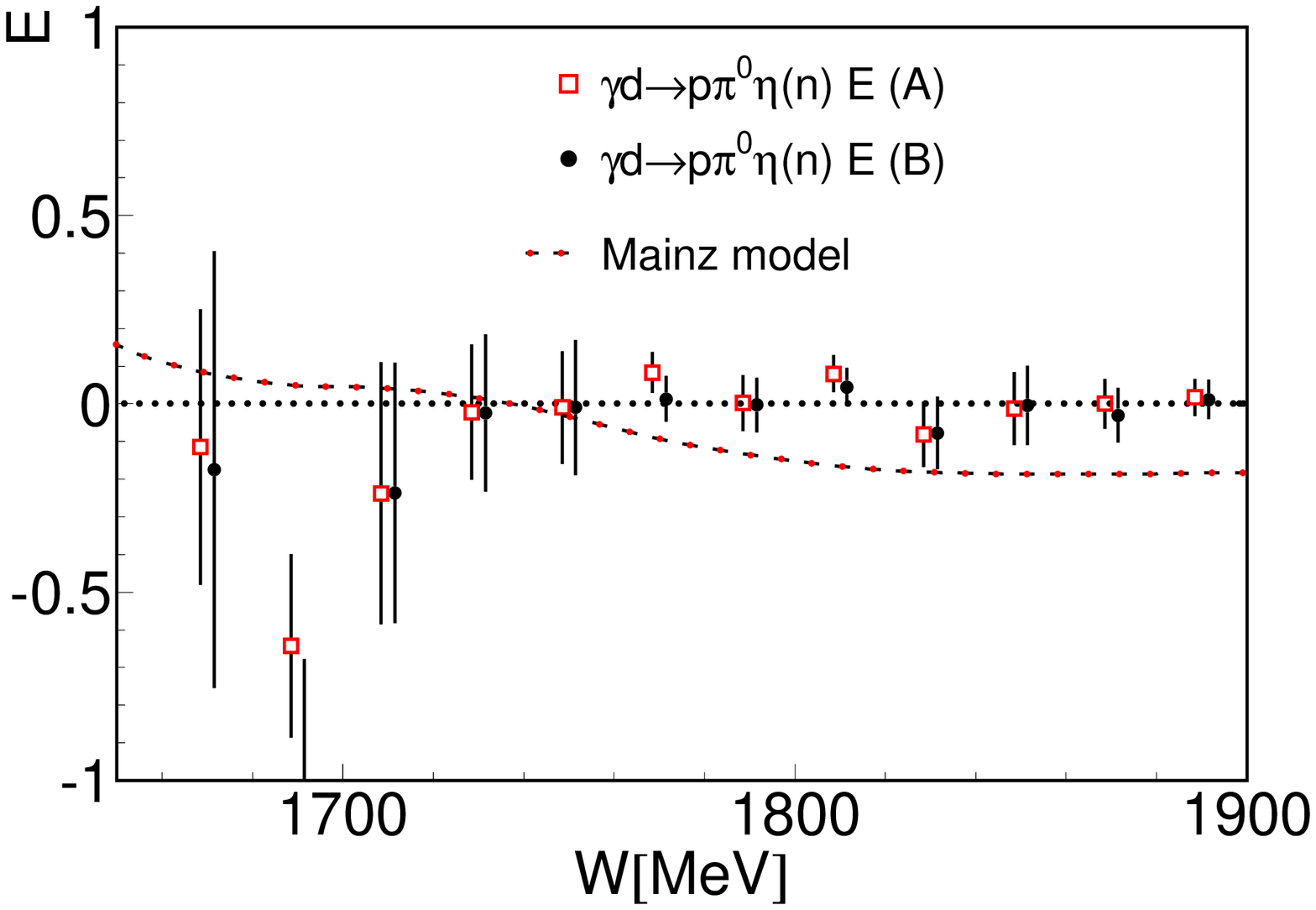}}
\resizebox{0.5\textwidth}{!}{\includegraphics{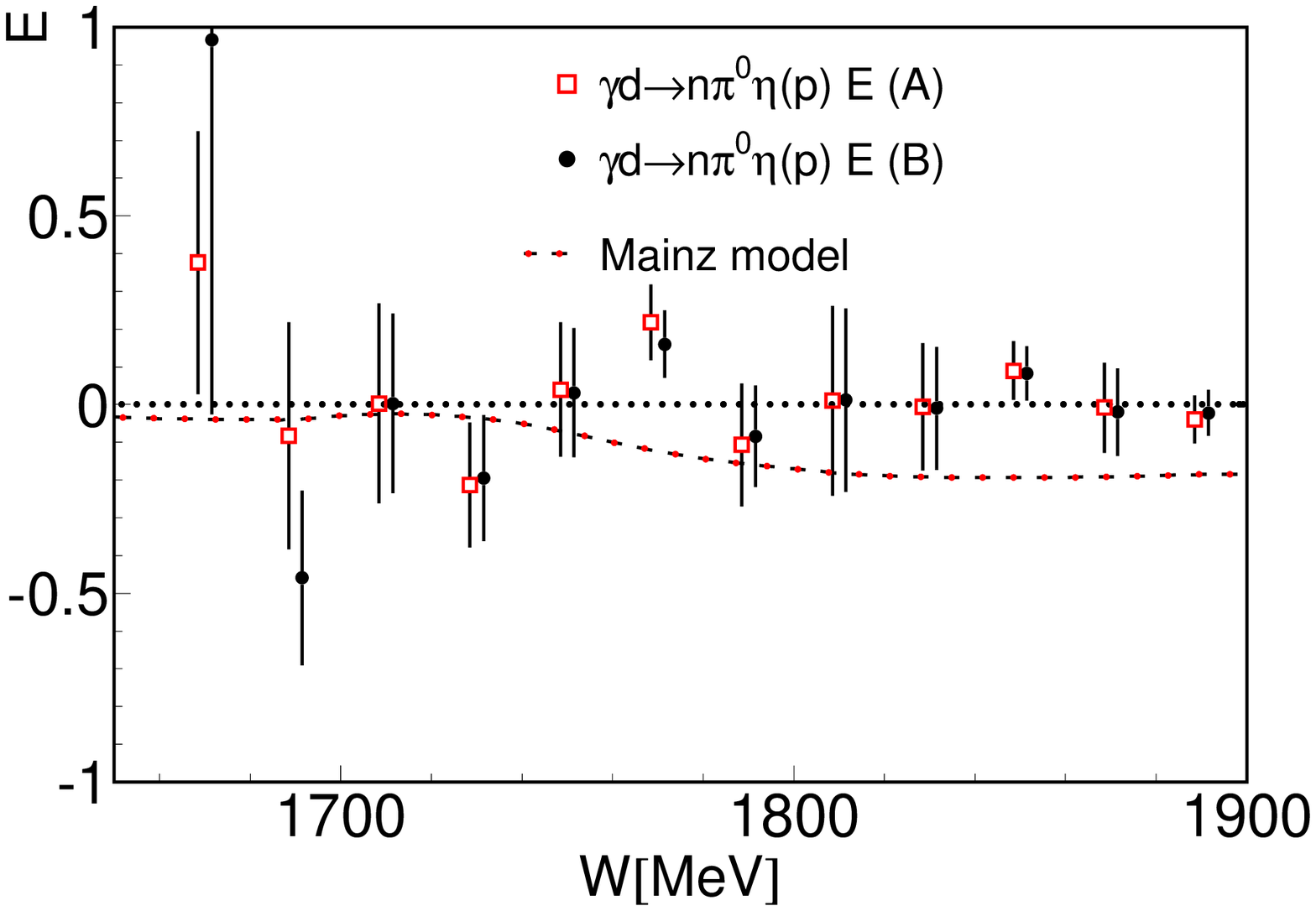}}}
\caption{Double polarization observable $E$. Left-hand side: quasi-free protons, 
right-hand side: quasi-free neutrons. 
Open (red) symbols: analysis (A) (normalization to unpolarized deuterium cross sections), 
closed (black) symbols: analysis (B) (subtraction of carbon background).
Data points shifted by $\pm$1.5~MeV from their nominal values for better readability of the figure.
Dashed lines: predictions from the Mainz model \cite{Sokhoyan_18}.}
\label{fig:E_pn}
\end{figure*}

The results for the double polarization observable $E$ (see Eq.~\ref{eq:e}) are shown in 
Fig.~\ref{fig:E_pn} as a function of the invariant mass $W$. The results from the two different 
analyses using either a normalization to the unpolarized cross section measured with a liquid deuterium 
target (analysis (A)) or the subtraction of the unpolarized carbon background in the denominator of 
Eq.~\ref{eq:e} (analysis (B)) are in good agreement, which demonstrates that systematic effects from 
normalizations and background subtraction are well under control. The statistical fluctuations of
both analyses are highly correlated. This was expected because the fluctuations are dominated by the 
almost vanishing numerator of the ratio in Eq.~\ref{eq:e}, which was identical for both analyses.

The result for the asymmetries is different from other reaction channels such as $\eta$ production 
\cite{Witthauer_16,Witthauer_17} and $\pi^0$ production \cite{Dieterle_17}. The asymmetry vanishes, 
within statistical uncertainties, over the full investigated energy range. The vanishing asymmetry is 
certainly not an instrumental effect because the same data set has already produced substantial 
asymmetries for production of $\eta$ mesons \cite{Witthauer_16,Witthauer_17}, single $\pi^0$ production 
\cite{Dieterle_17}, and production of pion pairs (not yet published).
This means that contributions to $\eta\pi^0$ production must be almost exactly balanced for the excitation 
of nucleon resonances via the $A_{1/2}$ and $A_{3/2}$ electromagnetic reaction amplitudes. This result 
was established for reactions off protons and off neutrons as expected for the primary excitation of $\Delta$ 
resonances. The most recent results for this reaction from an analysis of cross section data and photon-helicity 
asymmetries (circularly polarized photon beam, unpolarized target) in the framework of the Mainz model  
have been published in Ref. \cite{Sokhoyan_18} (the basis of this model is discussed in \cite{Fix_10}). 
The model predictions for the E asymmetry are plotted in Fig.~\ref{fig:E_pn}. In the region of the 
strongly contributing $\Delta(1700)3/2^-$ resonance agreement is reasonable within the relatively large 
statistical uncertainties of the experimental data.
At larger invariant masses, the predictions favor a negative asymmetry and deviate systematically
from the measured vanishing asymmetries. This deviation is even more apparent in the comparison of the
helicity-dependent cross sections in Fig.~\ref{fig:Heli_pn}.

\begin{figure}[!htb]
\centerline{
\resizebox{0.25\textwidth}{!}{\includegraphics{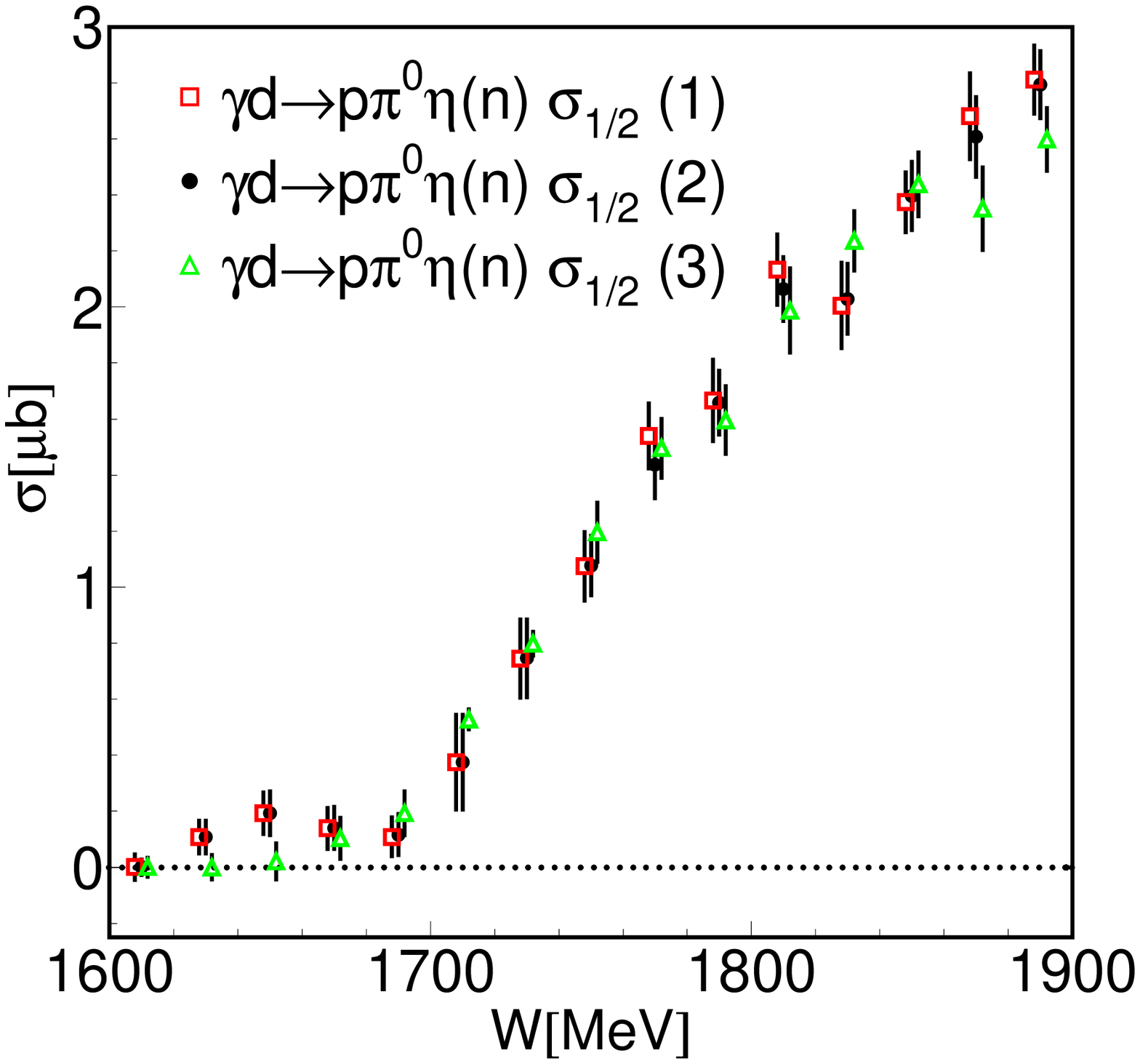}}
\resizebox{0.25\textwidth}{!}{\includegraphics{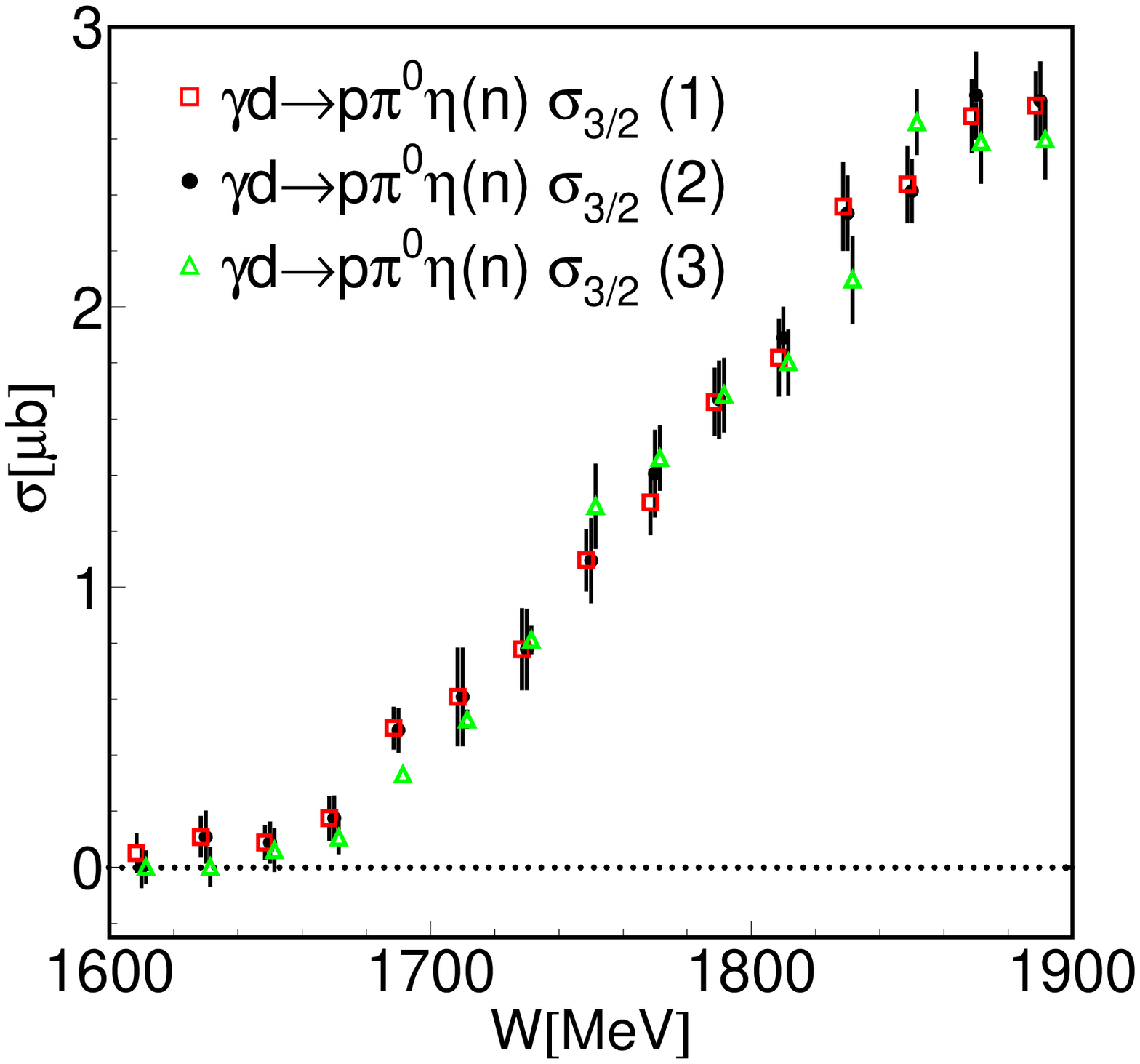}}}
\centerline{
\resizebox{0.25\textwidth}{!}{\includegraphics{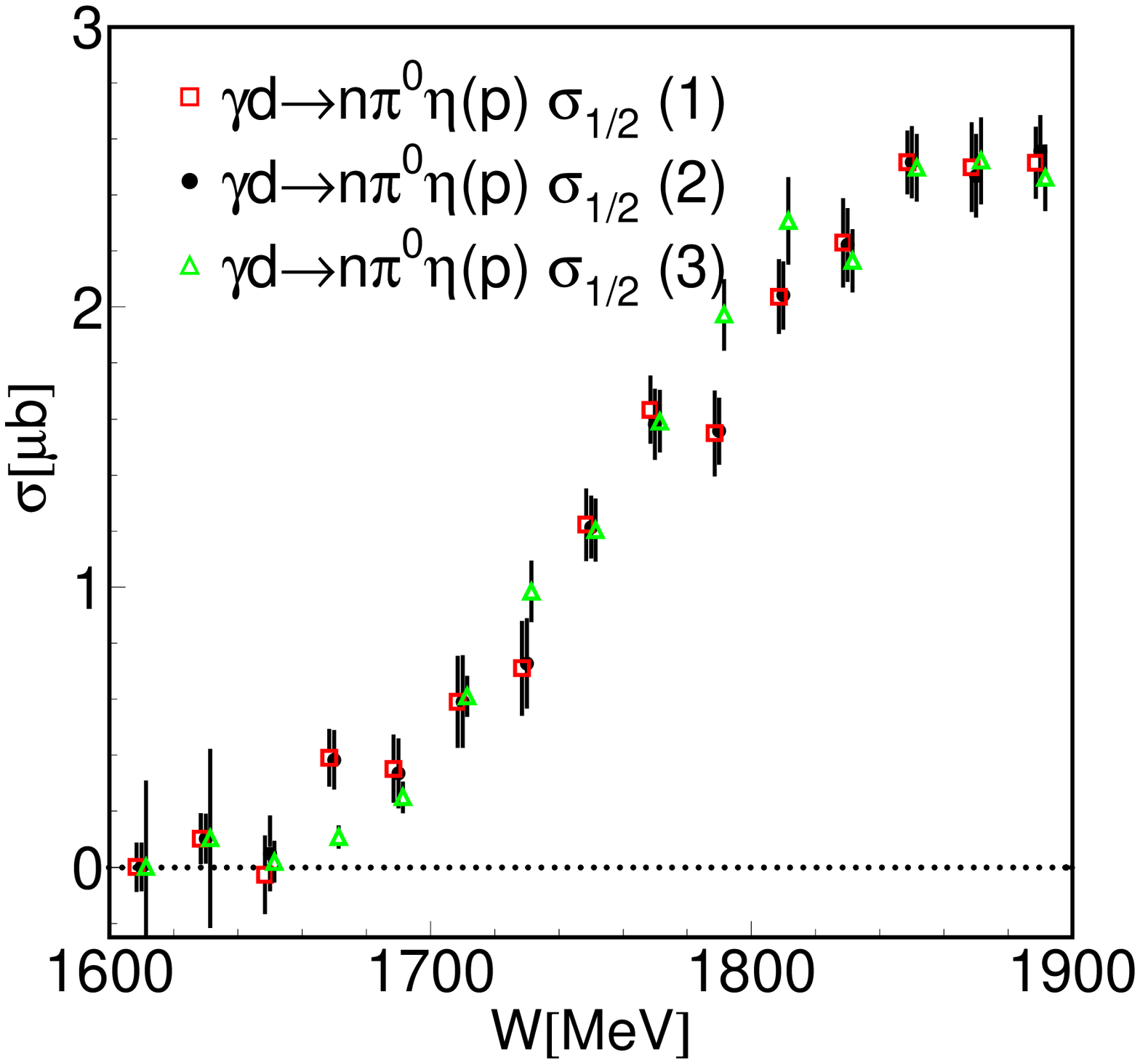}}
\resizebox{0.25\textwidth}{!}{\includegraphics{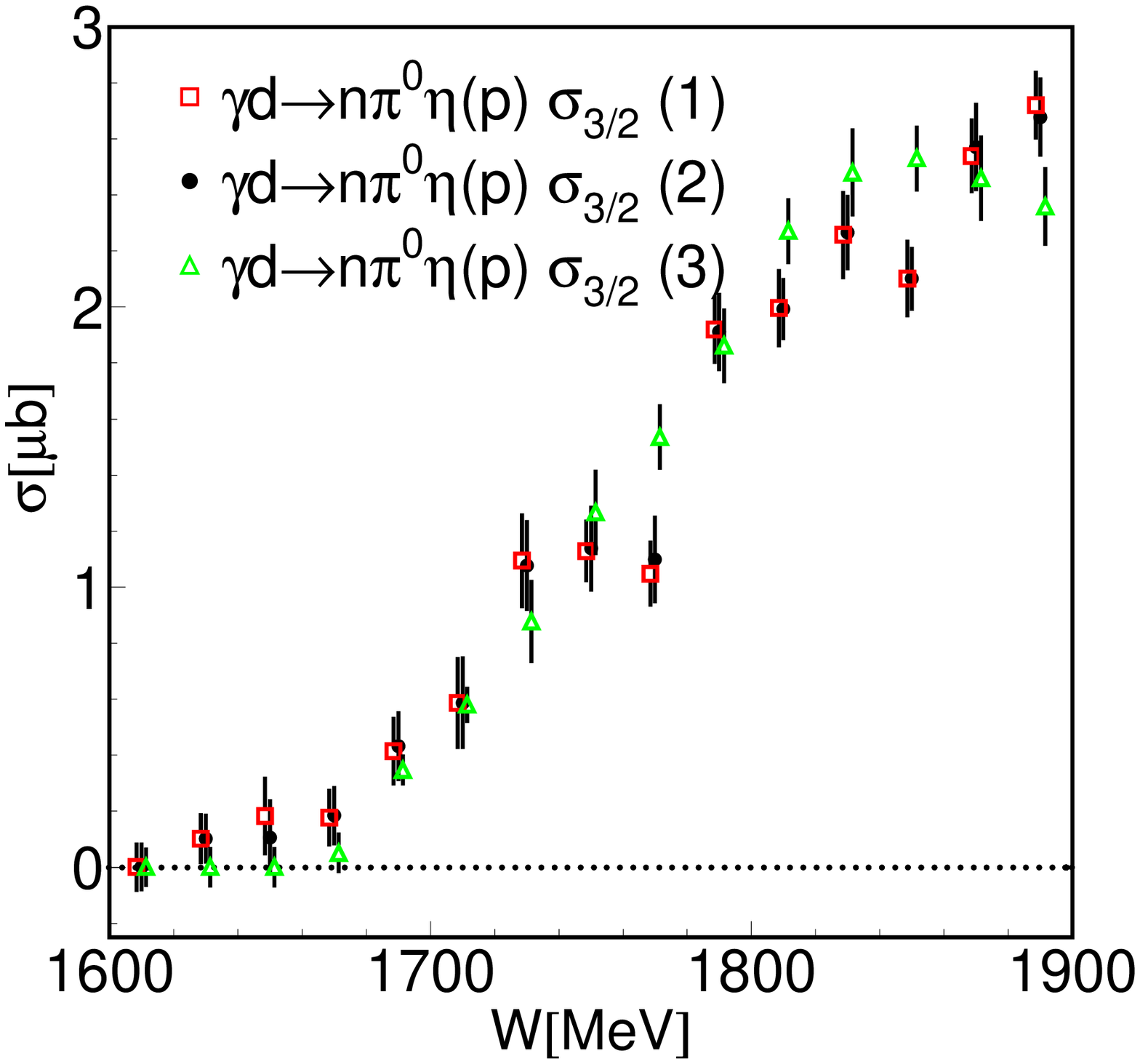}}}
\caption{Helicity-dependent cross sections $\sigma_{1/2}$ (left-hand side) and $\sigma_{3/2}$ 
(right-hand side) for quasi-free protons (upper row) and quasi-free neutrons (bottom row) for 
the three different analysis methods. 
version (1): (red squares): $E$ from analysis (A), $\sigma_0$ from unpolarized deuterium cross section,
version (2): (black filled dots): $E$ from analysis (B), $\sigma_0$ from unpolarized deuterium cross section,
version (3): (green triangles): $E$ from analysis (B), $\sigma_0$ also from butanol target with carbon subtraction.
Data points for (2) at nominal values, points for (1),(3) shifted by $\pm$1.5~MeV.}
\label{fig:all_heli}
\end{figure}

\begin{figure*}[!thb]
\centerline{
\resizebox{0.47\textwidth}{!}{\includegraphics{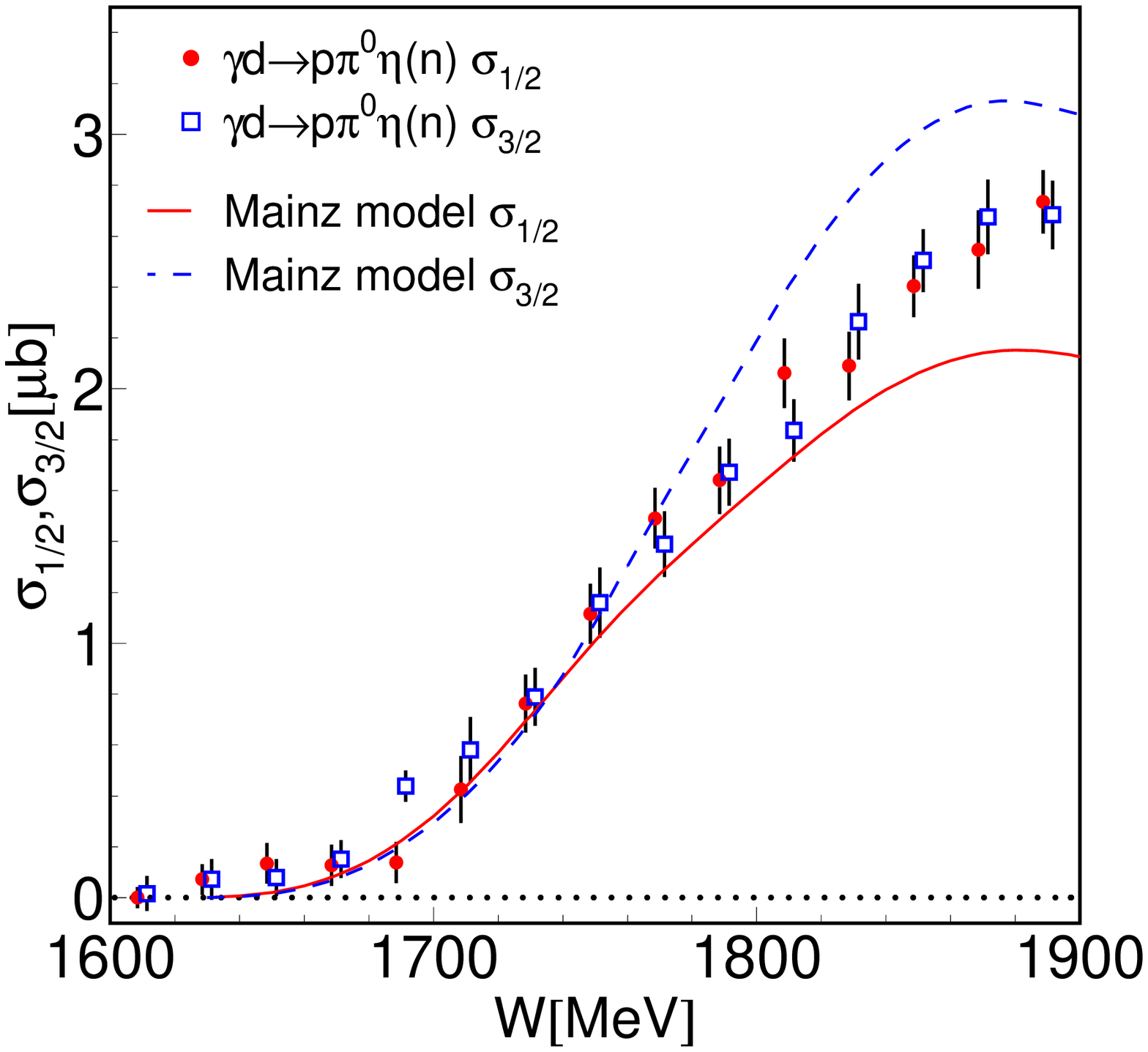}}\hspace*{1cm}
\resizebox{0.47\textwidth}{!}{\includegraphics{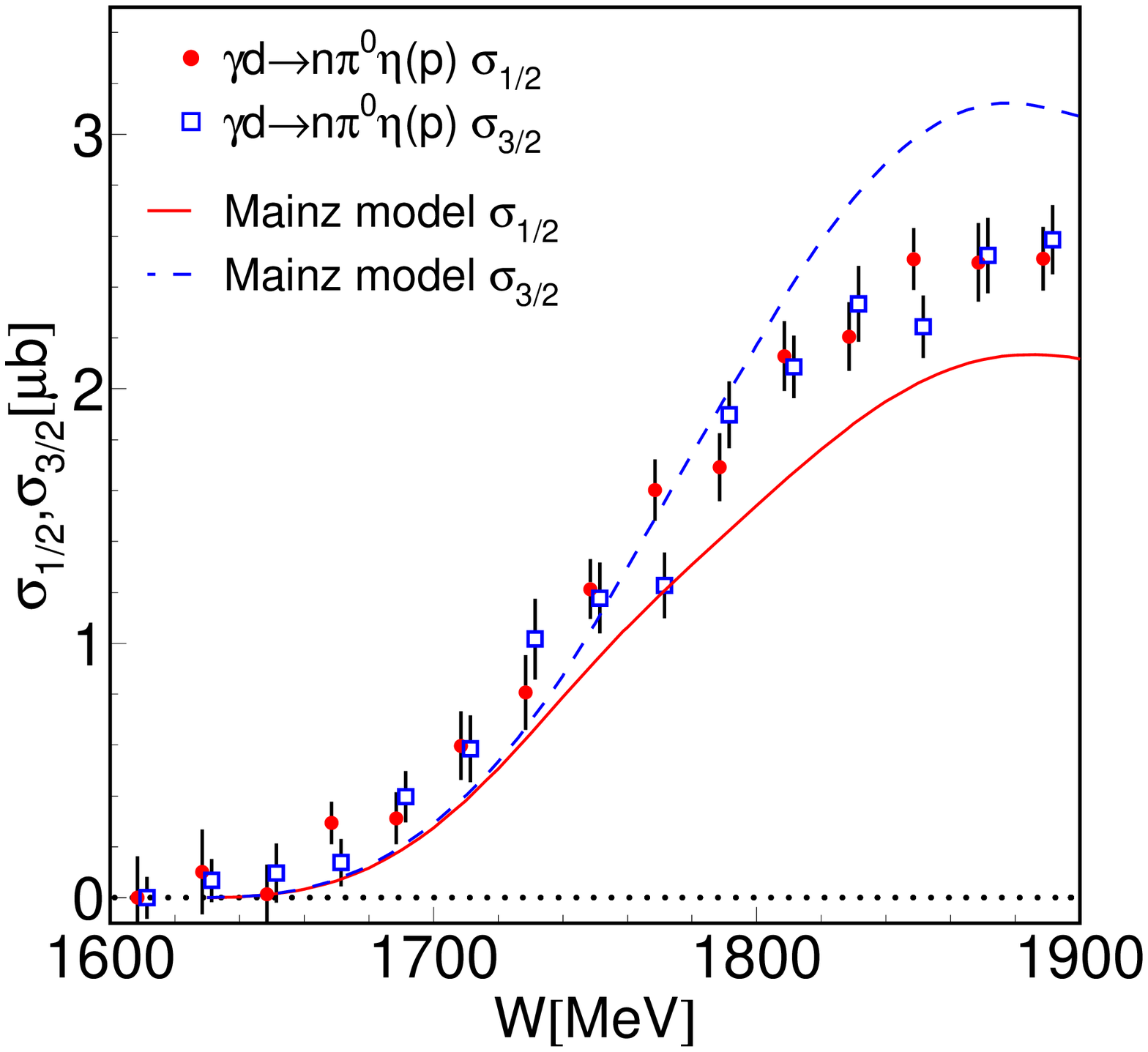}}}
\caption{Helicity-dependent cross sections $\sigma_{1/2}$ (filled red dots) and $\sigma_{3/2}$ 
(open blue squares) as function of total cm energy $W$. The results from the three different analyses 
(see Fig.~\ref{fig:all_heli}) have been averaged. 
Left-hand side: coincidence with recoil protons, right-hand side: coincidence with recoil neutrons.
Data points shifted by $\pm$1.5~MeV from their nominal values.
Solid (red) lines and dashed (blue) lines: results from the Mainz model \cite{Sokhoyan_18}. 
All model curves are scaled down by a common factor of 1.3 to account for FSI effects.}
\label{fig:Heli_pn}
\end{figure*}

The helicity-dependent cross sections from the three different analyses for coincident recoil protons and 
neutrons are summarized in Fig.~\ref{fig:all_heli}. The three analyses agree within statistical 
uncertainties, which indicates that there are no serious systematic effects, either from the use of the 
unpolarized cross section measured with a liquid deuterium target, or from the carbon subtraction. 
The agreement of analysis (3) with the other two results means that not only the asymmetries, but also
the absolute cross section, can be extracted from the carbon-subtracted butanol data.

The results from the three analyses were averaged for the final results of the helicity-dependent cross 
sections, which are shown in Fig.~\ref{fig:Heli_pn}. Since the statistical fluctuations of the three 
extractions are strongly correlated due to the correlation of the numerator for analysis (A) and (B) of 
the asymmetry $E$ and the use of $\sigma_0$ from the liquid deuterium target for analysis (1) and (2), the 
statistical uncertainties were combined linearly, rather than quadratically. The main result is that for 
quasi-free protons, as well as for quasi-free neutrons, the two helicity cross sections $\sigma_{1/2}$ and 
$\sigma_{3/2}$ agree within statistical uncertainties. Also, as expected from the results in 
\cite{Kaeser_15}, the results for the neutron and the proton are almost identical in magnitude. One should, 
however, note that the results for the total unpolarized cross section for the quasi-free proton are 
affected by FSI as discussed in \cite{Kaeser_15,Kaeser_16}. Compared to reactions on the free proton, 
cross sections are lower by approximately 30\%. The results are compared to the model predictions
from \cite{Sokhoyan_18}. All model results have been scaled down by a factor of 1.3 to account for the 
reduction of the quasi-free cross sections in absolute magnitude with respect to free-nucleon cross sections
as observed in \cite{Kaeser_16}. The comparison emphasizes what can already be seen in Fig.~\ref{fig:E_pn}.
In the energy range around 1700~MeV, dominated by the $\Delta(1700)3/2^-$ resonance, experimental data
and model predictions agree in so far as the two helicity dependent cross sections are equal within 
uncertainties (Ref.~\cite{Sokhoyan_18} quotes an $A^{3/2}/A^{1/2}$ ratio of 0.8). Small deviations between 
experimental data and model results on an absolute scale may be due to the rough 30\% correction of FSI effects
which may also have some energy dependence. However, at invariant masses above 1750~MeV the model predicts
a clear dominance of the $\sigma_{3/2}$ part of the cross section, which is not seen in the measured data.
In the model fit, this arises from large $A^{3/2}/A^{1/2}$ ratios for the $\Delta(1920)3/2^+$ and the  
$\Delta(1940)3/2^-$ states. These ratios were smaller in the original version of this model \cite{Fix_10}
(see Table I in \cite{Sokhoyan_18}) and they were much smaller in the BnGa model \cite{Gutz_14} but increased 
in the more recent Mainz fit of several differential cross sections \cite{Sokhoyan_18}.   
    
\section{Summary and Conclusions}
The double-polarization observable $E$ and the helicity-dependent cross sections $\sigma_{1/2}$ 
and $\sigma_{3/2}$ were measured for photoproduction of $\pi^{0}\eta$ pairs from quasi-free protons and 
neutrons. As already reported in \cite{Kaeser_15,Kaeser_16}, the reactions off protons and neutrons have 
almost exactly identical cross sections. Compared on an absolute scale to the free-proton cross sections, 
they are, however, significantly reduced due to FSI effects. The first measurement of the helicity 
dependence shows in addition that for both target nucleons the asymmetry $E$ is consistent with zero. 
This means that contributions from the two helicity states must be exactly balanced over the full energy 
range explored. The most natural explanation for both observations is that this reaction is dominated by 
the excitation of one (or few) $\Delta$ resonances decaying via $\eta$ emission to the $\Delta$(1232) 
with subsequent pion decay to the nucleon ground state and that the electromagnetic excitation
amplitudes of the primarily excited $\Delta$ states are nearly identical for both helicity states.
The new data will certainly much constrain future partial wave analyses in this energy range.
Comparison to existing model predictions shows that the ratio of the so far poorly known $A_{1/2}/A_{3/2}$ 
helicity couplings for the higher lying $\Delta$ states must be almost certainly revised. 
The RPP \cite{PDG_16} values for this parameters have still large uncertainties, which probably 
cover the range of needed modifications, but the present data will constrain them much tighter.  

\newpage
\noindent{\bf Acknowledgments}\\

We wish to acknowledge the outstanding support of the accelerator group and operators of MAMI.
We thank L.~Tiator for very useful discussions. 
This work was supported by Schweizerischer Nationalfonds (200020-156983, 132799, 121781, 117601),
Deutsche For\-schungs\-ge\-mein\-schaft (SFB 443, SFB 1044, SFB/TR16), the INFN-Italy,
the European Community-Research Infrastructure Activity under FP7 programme (Hadron Physics,
grant agreement No. 227431),
the UK Science and Technology Facilities Council (ST/J000175/1, ST/G008604/1, ST/G008582/1,ST/J00006X/1, and
ST/L00478X/1),
the Natural Sciences and Engineering Research Council (NSERC, FRN: SAPPJ-2015-00023), Canada. This material
is based upon work also supported by the U.S. Department of Energy, Office of Science, Office of Nuclear
Physics Research Division, under Award Numbers DE-FG02-99-ER41110, DE-FG02-88ER40415, DE-FG02-01-ER41194,
and DE-SC0014323 and by the National Science Foundation, under Grant Nos. PHY-1039130 and IIA-1358175.

\end{document}